\documentclass[11pt,preprint]{aastex}
\usepackage{apjfonts}

\def\gax{\mathrel{\raise.3ex\hbox{$>$}\mkern-14mu\lower0.6ex\hbox{$\sim$}}}
\def\lax{\mathrel{\raise.3ex\hbox{$<$}\mkern-14mu\lower0.6ex\hbox{$\sim$}}}
\def\gtorder{\mathrel{\raise.3ex\hbox{$>$}\mkern-14mu
             \lower0.6ex\hbox{$\sim$}}}
\def\ltorder{\mathrel{\raise.3ex\hbox{$<$}\mkern-14mu
             \lower0.6ex\hbox{$\sim$}}}

\begin{document}

\title{The Astrophysical Implications of Dust Formation \\ During The Eruptions of Hot, Massive Stars }

\author{
   C.~S. Kochanek$^{1,2}$ 
  }

\altaffiltext{1}{Department of Astronomy, The Ohio State University, 140 West 18th Avenue, Columbus OH 43210}
\altaffiltext{2}{Center for Cosmology and AstroParticle Physics, The Ohio State University, 191 W. Woodruff Avenue, Columbus OH 43210}

\begin{abstract}
Dust formation in the winds of hot stars is inextricably linked to the classic 
eruptive state of luminous blue variables (LBVs) because it requires very high mass loss
rates, $\dot{M}\gtorder10^{-2.5}M_\odot$/year, for grains to grow and for
the non-dust optical depth of the wind to shield the dust formation region from the 
true stellar photosphere.  Thus, dusty shells around hot stars trace the history of
``great'' eruptions, and the statistics of such shells in the Galaxy indicate that 
these eruptions are likely the dominant mass loss mechanism for evolved, $M_{ZAMS}\gtorder40M_\odot$
stars.  Dust formation at such high $\dot{M}$ also explains why very large grains 
($a_{max}\gtorder1\mu$m) are frequently found in these shells, since $a_{max}\propto\dot{M}$.
The statistics of these shells (numbers, ages, masses, and grain properties
such as $a_{max}$) provide an archaeological record of this
mass loss process.  In particular, the velocities $v_{shell}$, transient
durations (where known) and ejected masses $M_{shell}$ of the Galactic shells and
the supernova ``impostors'' proposed as their extragalactic counterparts 
are very different.  While much of the difference is a selection 
effect created by shell lifetimes $\propto(v_{shell}\sqrt{M_{shell}})^{-1}$,  
more complete Galactic and extragalactic surveys are needed to demonstrate that
the two phenomena share a common origin given that their observed properties
are essentially disjoint.  If even small fractions ($1\%$) of SNe show interactions 
with such dense shells of ejecta, as is currently believed, then the driving 
mechanism of the eruptions must be associated with the very final phases
of stellar evolution, suggestive of some underlying nuclear burning instability.
\end{abstract}


\section{Introduction}
\label{sec:introduction}

The winds of hot stars do not form dust, as illustrated by the single uses
of the word ``dust'' in the reviews of these winds by \cite{Kudritzki2000} 
and \cite{Puls2008}.
The exceptions which prove the rule are the rare, dust forming WC
stars, all of which appear to be binaries where dust forms due to
the collision of the two stellar winds (see the review by \citealt{Crowther2007}),
and the relatively rare B[e] stars where dust is believed to form in a disk/dense equatorial
wind surrounding the star (see the review by \citealt{Waters1998}).  
Dust formation is inhibited by the low particle densities and the harsh
ultraviolet (UV) environment (e.g. \citealt{Cherchneff1995}).  Cool
star winds readily form dust, and there are extensive studies of dust formation
in such environments (see the review by \citealt{Willson2000}).
 
Yet it is clear the hot, massive stars can episodically form enormous
quantities of dust and that this is related to the eruptions of 
luminous blue variables (LBVs, see the reviews by \citealt{Humphreys1994}, \citealt{Vink2009}).
LBVs are observed in three states: a quiescent, hot ($T_*>15000$~K) state,
a cooler ($T_*\simeq 7000$~K) eruptive or S Doradus state of roughly the same 
bolometric luminosity but enhanced mass loss rates 
($\dot{M} \sim 10^{-4}$ to $10^{-5}M_\odot$/year),
and a similarly cool, ``great'' eruptive state of significantly higher 
bolometric luminosity and enormously enhanced
mass loss rates ($\dot{M} \gtorder 10^{-2}$--$10^{-3} M_\odot$/year). 
We will call these the hot (or quiescent), cool (or S Doradus), 
and (great) eruptive states, and we will refer to the ejected material 
from an eruption as a shell since the low duty cycles of eruptions 
produce relatively thin dusty shells of ejecta. 
In their (great) eruptive state, these stars can expel enormous amounts
of material under conditions favorable to the growth of dust grains,
as illustrated by the massive ($\sim 10M_\odot$, \citealt{Smith2003}),
optically thick, dusty shell surrounding $\eta$ Carinae (see the reviews by
\citealt{Davidson1997}, \citealt{Smith2009}).  Indeed, it
is likely that such phases represent the bulk of the mass loss
from higher mass stars ($M\gtorder 40M_\odot$) because normal
winds are inadequate to the task (\citealt{Humphreys1984},
\citealt{Smith2006a}).  The
recent discovery of many $24\mu$m shells surrounding hot stars
by \cite{Wachter2010} and \cite{Gvaramadze2010} further suggests 
that the phenomenon is more common than previously thought.  

Table~1 summarizes the properties of Galactic LBVs and 
LBV candidates mainly drawn from \cite{Humphreys1994} and 
\cite{Smith2006a}.  All the stars have substantial, 
$\dot{M}\sim 10^{-5}M_\odot$/year,
relatively fast, $v_\infty \simeq 200$~km/s, present day winds
that are not forming dust, as expected for hot stellar winds.
However, all but one system is surrounded by a relatively
massive, $M_{shell} \sim M_\odot$, slowly expanding, 
$v_{shell} \sim 100$~km/s, shell of dusty material, 
and in at least four cases, models of the shell appear
to require surprisingly large maximum grain sizes,
$a_{max}\gtorder 1\mu$m.  We focused on these sources
because most of these ancillary properties have been
measured.  Most of the mass estimates are based on 
assuming a dust-to-gas ratio $X_d=0.01$ and so could
be underestimates.

While Table~1 is certainly incomplete and subject to many 
selection effects, that it contains 13 objects means that
shell ejections are an important or even dominant mass loss 
process for massive stars, as has been previously suggested 
in order to compensate for the steady downward revisions of the
mass loss rates in normal, hot stellar winds (e.g., 
\citealt{Humphreys1984}, \citealt{Smith2006a}).
We can quantify this by estimating the number of dusty
shells that should exist in the Galaxy given the rate of
Galactic supernovae, $r_{SN}$.  For simplicity, we use
a Salpeter initial mass function (IMF) and assume that supernovae
arise from stars with initial masses $M_*$ in the range 
$M_{SN}\simeq 8M_\odot \ltorder M_* \ltorder M_{up}$, where
for now we will let $M_{up} \rightarrow \infty$.  If
eruptions occur in stars with $M_* > M_{erupt}$ and there are an
average of $N_{erupt}$ occurrences per star, then the eruption
rate $r_{erupt}$ is of order 
\begin{equation}
      r_{erupt} \simeq 0.1 
           \left( { 40 M_\odot \over M_{erupt} } \right)^{1.35}N_{erupt} r_{SN}.
\end{equation}     
The optical depth of a dusty shell of mass $M_{shell}$ with visual opacity
$\kappa_V \simeq 100$~cm$^2$/g expanding at velocity $v_{shell}$ is $\tau_V=M_{shell}\kappa_V /4\pi v_{shell}^2 t^2$
so the shell will be detectable for of order
\begin{equation}
     t_{shell} \simeq \left({ M_{shell} \kappa_V  \over 4 \pi v_{shell}^2 \tau_V } \right)^{1/2}
          \simeq 5700 \left( { 0.1 \over \tau_V } \right)^{1/2} \left( { M_{shell} \over 10 M_\odot } \right)^{1/2}
            \left( { 70~\hbox{km/s} \over v_{shell} } \right)~\hbox{years}.
         \label{eqn:tshell}
\end{equation}
Most shells should be seen near their maximum size, $v_{shell} t_{shell} \simeq 0.4$~pc, which
is typical of the examples in Table~1.  
The total fraction of the stellar 
luminosity reradiated in the mid-IR is larger than $\tau_V$ because of the increased dust
opacity in the UV.  The expected number of Galactic shells is the product of the rate and
the lifetime,
\begin{equation}
    N_{shell} = r_{erupt} t_{shell} = 6 N_{erupt} \left( { 0.1 \over \tau_V }\right)^{1/2} \left( { r_{SN} \over \hbox{century}^{-1} } \right)
            \left( { 40 M_\odot \over M_{erupt} } \right)^{1.35}
            \left( { M_{shell} \over 10 M_\odot } \right)^{1/2} \left( { 70 \hbox{km/s} \over v_{shell} } \right).
\end{equation}
As we see from Table~1, there are at least $N_{shell} \sim 10$ LBV stars surrounded
by massive, dusty shells in the Galaxy, which means that the number of eruptions per star is
\begin{equation}
        N_{erupt} \simeq 2 \left( { N_{shell} \over 10 } \right) \left( {\tau_V \over 0.1} \right)^{1/2} 
             \left( { \hbox{century}^{-1} \over r_{SN}} \right)
            \left( { M_{erupt} \over 40 M_\odot  } \right)^{1.35}
            \left( { 10 M_\odot \over  M_{shell}} \right)^{1/2} \left( { v_{shell} \over 70 \hbox{km/s} } \right).
\end{equation}
and the amount of ejected mass per star due to the eruptions is $M_{tot} = N_{erupt} M_{shell} \simeq 15 M_\odot$.  
A ``normal'' wind from a hot star with  $\dot{M}_{normal} \simeq 10^{-5} M_\odot$/year 
would have to operate continuously for over $10^6$~years to equal the typical eruptive mass loss implied
by the existence of even the well-studied Galactic shells.   That $N_{erupt} > 1$ is also consistent with the existence of multiple shells around some
of the Galactic examples (e.g. G72.29+0.46, \citealt{Jimenez2010}).  Note, however, that the overall duty
cycle of the shell phase is low, since $N_{erupt} t_{shell} \simeq 10^4$~years as compared
to post-main-sequence lifetimes of order $10^6$~years.  These estimates are broadly consistent
with earlier estimates (e.g. \citealt{Humphreys1994}, \citealt{Lamers1989}) but based on a
different approach.
  
Given such a large contribution to the mass loss history of massive stars, we
need to understand the relationship between mass loss and
dust formation around hot stars.  We consider a parcel of fluid ejected
in a wind of mass loss rate $\dot{M}$ and velocity $v_w$ ejected from
a star of luminosity $L_*$ and temperature $T_*$ and examine the
conditions under which dust can form in \S\ref{sec:dform}.
Not surprisingly, the key variable is the mass loss rate.  First, for 
stellar winds with velocities of order the escape velocities of massive
hot stars, very high mass loss rates are needed for particle growth.
Second, the dust formation region must be shielded from the hot stellar
photosphere, which these high density winds can achieve by forming a 
pseudo-photosphere in the wind with a characteristic temperature of roughly $7000$~K.
{\it Dust formation around hot blue stars is necessarily
tied to very high mass loss rates, the classic LBV eruptive state and
the formation of shells of ejecta.}
In \S\ref{sec:discussion} we discuss some implications of this 
model for dust formation, stellar evolution and supernovae.

\section{The Physics of Dust Formation in Stellar Transient Ejecta}
\label{sec:dform}

\begin{figure}[p]
\centerline{\includegraphics[width=5.5in]{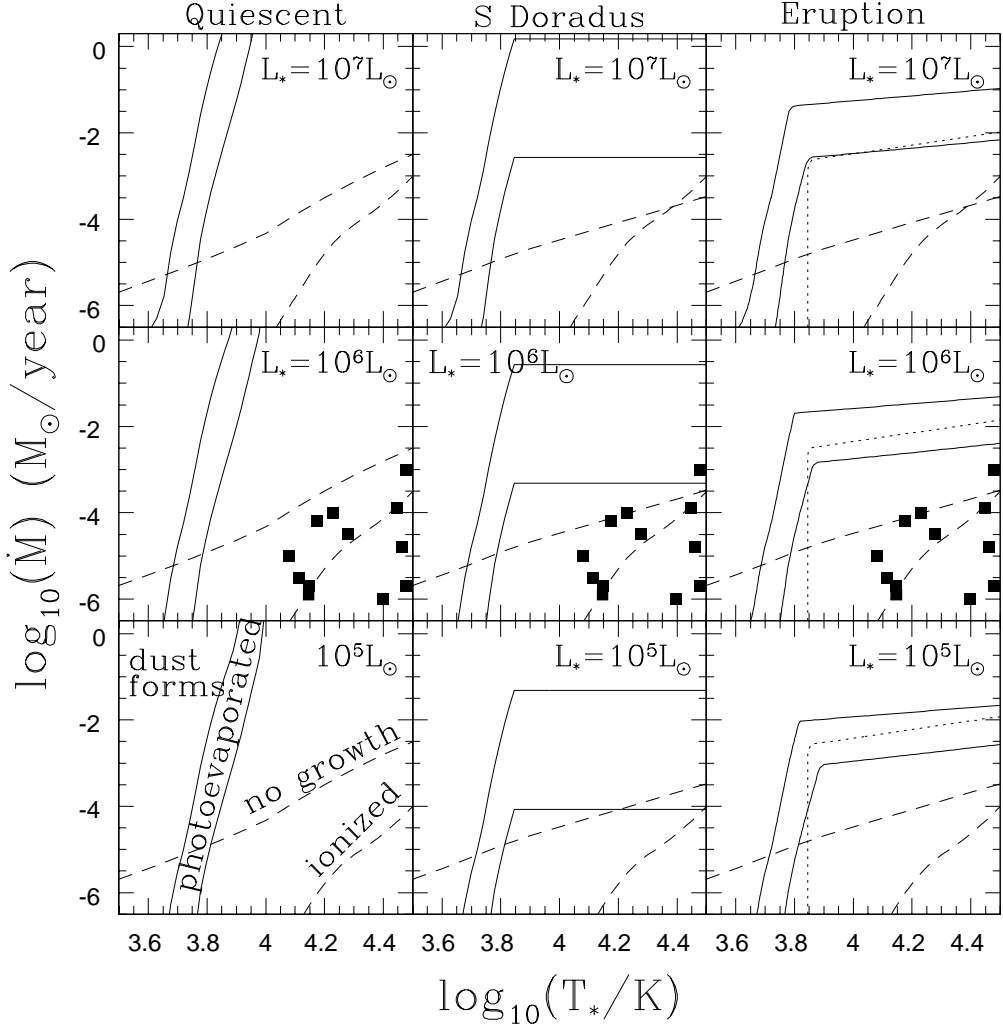}}
\caption{ 
  Minimum mass loss rates $\dot{M}$ for silicate dust formation as a function of stellar 
  temperature $T_*$.  The results for graphitic dusts are very similar.
  The lower left panel labels the regions.  Dust can form
  in the region labeled ``dust forms'', above the photoionization (``ionized''),
  growth (``no growth'') and photoevaporation (``photoevaporated'') limits. 
  The photoevaporation limits are for minimum photon energies of $E_0=7.5$ (stronger)
  and $10$~eV (weaker).  The left, middle and right columns are for the
  three different assumptions about the apparent photospheric temperature.
  In the quiescent state (left) the photospheric temperature is the stellar temperature,
  $T_{phot}=T_*$, in the S Doradus state (middle) the photospheric temperature
  of $T_{phot}=\hbox{min}\left(T_*,7000~\hbox{K}\right)$ is assumed to be determined by
  an expansion of the stellar photosphere that is uncorrelated with the wind, and in the
  eruptive state (right) the apparent temperature is determined by the non-dust optical
  depth of the wind.  The dotted contour in this column is the contour
  where $T_{phot}=7000$~K.  The upper, middle and lower panels show the changes
  for stellar luminosities of $L_*\equiv L_{phot} =10^7$, $10^6$ and $10^5 L_\odot$, respectively.
  The filled points are the present day properties of the systems from Table~1. 
  Other parameters are set to $M_*=20 M_\odot$, $v_c=1$~km/s, $X_g=0.005$,
  $T_d=1500$~K, $N=7$, $\hat{\rho}=3.8$ and $m_0=20m_p$.
  }
\label{fig:dform1}
\end{figure}

We consider the formation of dust grains of radius $a$ comprised of $N$
atoms of average mass $m_0$ where $4\pi a^3 \rho_{bulk}/3 = N m_0$ and 
$\rho_{bulk}$ is the bulk density of the grain.  We will use $\rho_{bulk}=2.2$~g/cm$^3$
($3.8$~g/cm$^3$) and $m_0=12 m_p$ ($20 m_p$) for graphitic (silicate) grains.
The smallest possible grain is the point where the inter-particle bond strengths
shift from strong molecular bonds to weaker intra-molecular interactions, and we will generally require $N\geq 7$ based
on
the number of atoms in Mg$_2$SiO$_4$.  There are many prior treatments of
dust formation in (generally cool) stellar winds (e.g. \citealt{Salpeter1977},   
\citealt{Draine1979}, \citealt{Gail1984}), novae (see the review by \citealt{Gehrz1988})
and supernovae (e.g. \citealt{Clayton1979}, \citealt{Dwek1988}, \citealt{Kozasa1991})
which contain most of the basic physical picture we use here.  For some standard
results in dust physics we will refer to \cite{Draine2011} as a reference source.
For dust to grow, the medium must be largely neutral, sufficiently cool for growth to occur, have a high enough
density for there to be an appreciable particle collision rate, and the grains
must grow faster than they can be photo-evaporated by ultraviolet (UV) photons. 
We will assume that the first stage of particle formation, nucleation to form the smallest
grains, simply occurs once the temperature is sufficiently low, and consider
only the subsequent collisional growth of the grains.  Making nucleation an
additional bottleneck to dust formation will only strengthen our conclusions.  

The dust forms in a (time varying) wind which we can characterize
by the mass loss rate $\dot{M}$, the wind velocity $v_w$, and the mass fraction of 
condensible species $X_g$.  Not all the condensible mass need condense onto
grains, so the ultimate mass fraction of dust $X_d \leq X_g$.  The wind is produced by
a star of luminosity $L_*$, photospheric radius $R_*$ and effective
temperature $T_*$ where $L_* = 4 \pi R_*^2 \sigma T_*^4$.  At the time the dust
is being formed, the star has luminosity $L_{phot}$, radius $R_{phot}$ and
temperature $T_{phot}$, with $L_{phot} = 4 \pi R_{phot}^2 \sigma T_{phot}^4$.
We consider dust formation in three physical states.  First, we have the
hot star in quiescence with $L_{phot}=L_*$ and $T_{phot}=T_*$.  Second,
we have the S Doradus state, where $L_{phot}=L_*$ and $T_{phot} \equiv 7000~\hbox{K} \ne T_*$.
Finally, we have the eruptive state where both $L_{phot} \ne L_*$ and $T_{phot} \ne T_*$
and we will use the optical depth of the wind to estimate $T_{phot}$.  In many cases we can
assume that the spectral energy distribution is simply a black body, 
but there are several areas where the differences between black bodies and
stellar photospheres are important because line opacities suppress the 
ultraviolet emission.  Where this is important, we will use the models
of \cite{Castelli2003}.  

For the physics of dust formation, the most relevant velocity is that near the dust 
formation radius since it sets the particle density that determines the growth
of the grains.  For hot stars, dust formation occurs sufficiently far from the star that the
wind acceleration should be largely complete and we can view the velocity as constant.   
We will scale the wind velocity by the surface escape velocity of the quiescent stars  
\begin{equation}
          v_e= \left( { 2GM_* \over R_* }\right)^{1/2}
             = 151 \left( { M_* \over 20 M_\odot }    \right)^{1/2} 
                   \left( { 10^6 L_\odot \over L_* }  \right)^{1/4}
                   \left( { T_* \over 10^4~\hbox{K} } \right)~\hbox{km/s},
         \label{eqn:vesc}
\end{equation}
as this is the typical asymptotic velocity scale of radiatively accelerated winds 
(see the reviews by \citealt{Kudritzki2000}, \citealt{Puls2008}).  Here
we have scaled the stellar mass to $M_*=20M_\odot$ under
the assumption that the stars have undergone significant mass loss by the time
of the eruption.  Since $v_e \propto T_*$, this
introduces a strong stellar temperature dependence to dust formation and growth. 
{\it In most of our results, 
 the wind velocity is always scaled by the escape velocity of the quiescent star!}
 Eqn.~\ref{eqn:vesc} yields an appropriate velocity scale for 
$\eta$~Carinae but may be somewhat high for many of the sources in Table~1.  
For the more massive examples in Table~1 the low shell expansion velocities
are almost certainly intrinsic because only a very
high density interstellar medium (ISM) can significantly slow the expansion of
these massive shells of material.\footnote{Given the shell masses it is very hard
to slow them down by large factors. Even slowing a $1 M_\odot$ shell
by a factor of two (from $140$ to $70$~km/s) within an expansion radius of $0.5$~pc requires an ISM density 
of order $10^2$~cm$^{-3}$. A dense ($\dot{M} \gtorder 10^{-5}M_\odot$/year), 
pre-existing slow wind from a red supergiant phase is a more promising means of 
having this much mass present, but the timing must be right and it still would 
not explain the slow multiple shell systems.  Slowing a massive ($10M_\odot$), 
fast ($500$~km/s) shell like that of $\eta$ Carinae down to $100$~km/s 
requires $40 M_\odot$ of material, and is essentially impossible.
  }
Moreover, some of the systems have multiple shells where the inner shells have low velocities but 
cannot be interacting with the ISM or an older slower wind 
(e.g. P~Cyg, \citealt{Meaburn1996}; G79.29$+$0.46 \citealt{Jimenez2010}).
Thus, it may be that some systems have asymptotic velocities significantly lower than
those from the surface of the quiescent hot star -- this can be approximated by
replacing $L_*$ and $T_*$ by $L_{phot}$ and $T_{phot}$, which will change the scaling 
of the wind velocity to use the escape velocity from the photosphere of the transient.

Luminous hot stars produce large numbers of ionizing photons, and dust cannot
form in such a hot, ionized medium.  For a pure hydrogen wind,
the wind can recombine if the rate of production of ionizing photons is less than
$Q_0 = \dot{M}^2 \alpha_B/4\pi v_w^2 m_p^2 R_*$ (e.g. \citealt{Fransson1982}).  The production of
ionizing photons is
\begin{equation}
      Q_0 = { L_* \over k T_* } F(G=x^{-1},E_1/kT_*,\infty)
\end{equation}
where $E_1=13.6$~eV, the dimensionless function is
\begin{equation}
    F(G,x_0,x_1) = 
     { \int_{x_0}^{x_1} G F_\nu d x  \over \int_0^\infty F_\nu d x }
     \rightarrow { 15 \over \pi^4 } \int_{x_0}^{x_1} { G x^3 dx \over \exp(x)-1 }, 
     \label{eqn:specfunc}
\end{equation}
where $x=h\nu/kT$ and the limit is that of a black body.   Thus, the minimum mass loss rate
for the wind to recombine is
\begin{eqnarray}
     \dot{M} &\gtorder &\left( { 8 \pi L_* G M_* m_p^2 F \over k T_* \alpha_B} \right)^{1/2} 
            \left({v_w \over v_e}\right) \nonumber \\
         &\simeq &2.1 \times 10^{-3} F^{1/2}
           \left( { 3 \times 10^{-13}~\hbox{cm}^3/\hbox{s} \over \alpha_B} 
                  { L_* \over 10^6 L_\odot } 
                  { 10^4~\hbox{K} \over T_* }
                  { M_* \over 20 M_\odot } \right)^{1/2} 
           \left( { v_w \over v_e } \right)~M_\odot/\hbox{year}.
\end{eqnarray}
For a black body, $F \simeq (15/\pi^4)(2+x(2+x)\exp(-x))$ where $x=158000/T_*$, making it
a small number unless the star is very hot ($F^{1/2} = 0.0024$ for a $T_*=10^4$~K black body), 
and the line blanketing of stellar atmospheres reduces it still further.  
Fig.~\ref{fig:dform1} shows this photoionization limit (``ionized'') on 
$\dot{M}$ for forming dust  based on the \cite{Castelli2003}  model atmospheres --
unless the star is very hot, ionization cannot prevent dust formation in dense winds.
Even then, the photoionization limit is only important for the quiescent star.

If the gas is relatively neutral, and so can carry out chemical reactions to form molecules,
the temperature must be low enough to aggregate the molecules into grains.
In general, the dust temperature is controlled by the radiation field because
collisional time scales are very much longer (see below).  We can divide the effects of
radiation into the equilibrium temperature and stochastic heating
of small grains by individual energetic photons which we discuss below.  
If the dust temperature is controlled by radiative heating, then dust 
can form once small grains will not be heated above the dust destruction temperature 
$T_d$.  If we consider only the mean temperature of the grains, then dust can form
outside radius (\citealt{Draine2011})
\begin{eqnarray} 
      R_{form} &= &\left( { L_{phot} Q_P(T_{phot},a_{min}) \over 16 \pi \sigma T_d^4 Q_P(T_d,a_{min}) }\right)^{1/2} =
           { R_* \over 2 }  \left( { L_{phot} \over L_* } \right)^{1/2}
              \left( { T_* \over T_d }\right)^2 \left( { Q_P(T_{phot},a_{min}) \over Q_P(T_d,a_{min}) } \right)^{1/2} \\
                  & = &5.2 \times 10^{14} 
                        \left( { L_{phot} \over 10^6 L_\odot } \right)^{1/2}
                         \left( { 1500~\hbox{K} \over T_d } \right)^2 
                         \left( { Q_P(T_{phot},a_{min}) \over Q_P(T_d,a_{min}) } \right)^{1/2}~\hbox{cm} \nonumber \\
             \label{eqn:rform}
\end{eqnarray}
where $Q_P(T,a_{min})$ is the Planck-averaged absorption efficiency for the smallest grains. To
simplify many subsequent results, we define
\begin{equation}
         Q_{rat} \equiv { Q_P(T_d,a_{min}) \over Q_P(T_{phot},a_{min}) },
\end{equation}
so $R_{form} = (R_*/2)(L_{phot}/L_*)^{1/2}(T_*/T_d)^2 Q_{rat}^{-1/2}$.  Note that the Planck factor for the star is
evaluated at the apparent photospheric temperature $T_{phot}$ which may not be the same as
the temperature at the stellar surface $T_*$.
Unless $T_d \simeq T_{phot}$, the corrections for the finite size of the star are unimportant and
for sufficiently small grains the result is independent of the grain size because
the $Q \propto a$  dependence cancels. If
we use the graphitic models of \cite{Draine1984} and $a_{min}=0.001\mu$m, the Planck average for
small graphitic dusts is approximately a power law
\begin{equation}
    Q_P(T)(\mu\hbox{m}/a_{min}) \simeq 0.42 \left( { T \over 1000~\hbox{K} } \right)^{3/2} 
\end{equation}
for $1000 < T < 50000$~K.  Transient photospheric temperatures are generally $T_{phot} \simeq 7000$~K,
so the formation radius for graphitic dusts is approximately
\begin{equation} 
          R_{form} \simeq 1.6 \times 10^{15} \left( { L_{phot} \over 10^6 L_\odot } \right)^{1/2}
                         \left( { 1500~\hbox{K} \over T_d } \right)^{7/2}~\hbox{cm}.
\end{equation}
The Planck averages for small silicate dusts cannot be reasonably approximated as a simple power
law, but a reasonable piecewise approximation is
\begin{equation}
     \log_{10}\left[ Q_P(T)(\mu\hbox{m}/a_{min})\right] \simeq -0.66 - 0.34 t_3 +1.15 t_3^2  \quad\hbox{for}\quad 10^3 < T < 10^4
\end{equation}
where $t_3=\log_{10}(T/1000~\hbox{K})$  and
\begin{equation}
     \log_{10}\left[ Q_P(T)(\mu\hbox{m}/a_{min})\right] \simeq 0.18 + 4.90 t_4 - 3.21 t_4^2 \quad\hbox{for}\quad 10^4 < T < 5 \times 10^4
\end{equation}
where $t_4=\log_{10}(T/10000~\hbox{K})$.  For temperatures in the range $1000 < T_d <2000$~K,
the Planck averages vary little, so for silicate dusts and $T_{phot} \simeq 7000$~K we find that
\begin{equation} 
      R_{form} \simeq 9.8 \times 10^{14}  \left( { L_{phot} \over 10^6 L_\odot } \right)^{1/2}
                   \left( { 1500~\hbox{K} \over T_d } \right)^2~\hbox{cm}.
\end{equation}
In general, including the Planck factors makes the formation radius roughly three times
larger than if they are ignored, and a reasonably general approximation is that
$R_{form} \simeq 10^{15} (L_{phot}/10^6L_\odot)^{1/2}$~cm.

\def\qbar#1{\left\langle Q_\lambda \left( #1 \right) \right\rangle}

If we assume that particle nucleation occurs rapidly once the ejecta are cool enough to form
dust, then the subsequent properties are limited by the growth of the dust particles. 
The collisional growth rate of a particle of radius $a$ is 
\begin{equation}
     { da \over dt } = { v_c X_g \dot{M} \over 16\pi v_w r^2 \rho_{bulk} }
    \label{eqn:dgrowth}
\end{equation}
where $v_c$ is an effective collisional velocity  (e.g. \citealt{Kwok1975}, \citealt{Deguchi1980}).
For thermal collisions,
accreting particles of mass $m_a$ at gas temperature $T$,
\begin{equation}
          v_c = \left( { 8 k T \over \pi m_a } \right)^{1/2}
              = 4.6 \left( { T \over 1000~\hbox{K} } \right)^{1/2} \left( { m_p \over m_a }\right)^{1/2}~\hbox{km/s}.
\end{equation}
This means that growth cannot proceed by coagulation of large particles if the particle
velocities are thermal because $m_a = N m_0$ means that $v_c \propto N^{-1/2}$ and the 
growth rate freezes out at very tiny grain sizes. In this case, growth must be dominated by the
accretion of monomers and very small clusters, so we can regard $m_a \simeq m_0$ as effectively
constant.  Coagulation will matter if $v_c$
is controlled by turbulent motions (e.g. \citealt{Voelk1980}) with the net effect that
particles can grow up to $4$ times faster than by monomer accretion.  
The gas presumably cools as it expands, so we will let $v_c=v_{c0} (R_{form}/R)^n$
where $R_{form}$ is the radius at which particle growth commences and $n=2/3$
if the cooling is dominated by adiabatic expansion and $n=1/4$ if it is controlled
by radiative heating at constant luminosity.  Other complications such as sticking
probabilities and exhausting the condensible species can be mimicked by adjusting
$v_{c0}$ or $n$.  
With these assumptions, we find that the particles grow to a maximum size of
\begin{eqnarray}
 a_{max} &= & { v_{cn} X_g \dot{M} \over 16 \pi \rho_{bulk} v_w^2 R_{form} } \\
              &\simeq & 2.5 \times 10^{-4} \left( { \dot{M} \hat{\rho}_{bulk}^{-1} \over 10^{-4} M_\odot/\hbox{year} } \right)
                   \left( { X_g \over 0.005 } \right)
                   \left( { v_{cn} \over \hbox{km/s}  } \right)
                   \left( { 500~\hbox{km/s} \over v_w  } \right)^2
                   \left( {  10^{15}~\hbox{cm} \over R_{form} } \right)~\mu\hbox{m} \nonumber \\
             &= &{ v_{cn} X_g \dot{M} \over 16 \pi G M_* \rho_{bulk} }
                    \left( { L_* \over L_{phot} } \right)^{1/2}
                    \left( { v_e \over v_w } \right)^2
                    \left( { T_d \over T_* } \right)^2  Q_{rat}^{1/2} \\
             &\simeq &5.3 \times 10^{-3} \left( { \dot{M} \hat{\rho}_{bulk}^{-1} Q_{rat}^{1/2}\over 10^{-4} M_\odot/\hbox{year} } \right)
                   \left( { X_g \over 0.005 } \right)
                   \left( { 20 M_\odot \over M_* } \right)
                   \left( { v_{cn} \over \hbox{km/s}  } \right)
                    \left( { L_* \over L_{phot} } \right)^{1/2}
                   \left( { v_e \over v_w } \right)^2
                   \left( { T_d \over 1500~\hbox{K}} { 10^4~\hbox{K} \over T_* } \right)^2~\mu\hbox{m} \nonumber 
    \label{eqn:amax}
\end{eqnarray}
where $v_{cn}=v_{c0}/(1+n)$ absorbs the effects of the cooling model on $a_{max}$ and 
$\hat{\rho}_{bulk}=\rho_{bulk}/\hbox{g/cm}^3$.  The grain size grows with radius as
\begin{equation}
       a=a_{max}\left( 1 - { R_{form}^{1+n} \over R^{1+n} } \right).
\end{equation}
Because the density is already dropping rapidly, reasonable assumptions about the
temperature scaling $n$ have little effect on the results.  Faster cooling leads
to smaller particles, but the full range from a constant temperature to adiabatic
cooling reduces $a_{max}$ by less than a factor of two, and the particles are close
to their final sizes by the time $R \simeq 2 R_{form}$.

If the mass loss rates are too low, then the particles cannot grow, which implies
there is a minimum mass loss rate for dust growth of
\begin{eqnarray}
   \dot{M} &> &{ 16 \pi G M_* \rho_{bulk} \over v_c X_g } 
            \left( { 3 N m_0 \over 4 \pi \rho_{bulk}} \right)^{1/3} 
            \left( { v_w \over v_e } \right)^2 
            \left( { L_{phot} \over L_* } \right)^{1/2} 
            \left( { T_* \over T_d} \right)^2  Q_{rat}^{-1/2} \label{eqn:mgrow1} \\
       &\gtorder & 3.2 \times 10^{-6} \left( { \hbox{km/s} \over v_c }\right)
             \left( { 0.005 \over X_g } \right)
             \left( { v_w \over v_e } \right)^2
            \left( { M_* \over 20 M_\odot } \right)
            \left( { L_{phot} \over L_* } \right)^{1/2} 
             \left( { 1500~\hbox{K} \over T_d } { T_* \over 10^4~\hbox{k} }\right)^2
              { \left( \hat{\rho}^{2} N \hat{m}_0\right)^{1/3}
                \over Q_{rat}^{1/2} }
             ~{M_\odot\over\hbox{year}} \nonumber \\
\end{eqnarray}
where we have phrased the limit in terms of the particle number $N$ rather
than the size $a$ and used $\hat{m}_0=m_0/12m_p$.  Fig.~\ref{fig:dform1} compares
these limits from particle growth (``no growth'') to those from photoionization.  
The limits from particle growth are always the more stringent.
When the wind velocity has $v_w \sim v_e$, it is very difficult
for hot stars to form dust because of the rapid increase in the wind
velocity with stellar temperature, $v_w \propto T_*$. 
We note that the limit at low temperatures 
appears high compared to typical AGB stars (e.g. \citealt{vanLoon2005}, 
\citealt{Matsuura2009}) primarily because the mass has been scaled to 
$M_*=20M_\odot$ and $\dot{M} \propto M_*$.

In the interstellar medium, the temperatures of the smallest dust grains are
stochastic because the absorption of individual photons can temporarily heat
the grains to temperatures far higher than the equilibrium temperature predicted
by the ambient radiation density (e.g. \citealt{Draine1985}, \citealt{Dwek1986}).  
This effect also plays a key role in the formation of dust by transients, but
seems not to have been generally considered outside estimates of dust formation
in the colliding wind environments of WC stars (e.g. \citealt{Cherchneff1995}).  
Under the assumption that the
ejecta must recombine in order to form dust, we are interested in photons
with energies  $E_0 < E  < E_1 \simeq 13.6$~eV since the hard UV photons
are absorbed near the base of the wind.  A small grain absorbing a soft
UV photon will be heated well above the equilibrium temperature and then
lose particles before radiatively cooling.  Grains cannot grow if this photoevaporation 
rate is faster than the collisional growth rate (\citealt{Draine1979b}).  We can estimate $E_0$ using the models of
\cite{Guhathakurta1989} for stochastic dust heating as the energy
at which the grain cooling time scale equals the time to lose
an atom from the grain.  This energy depends crucially on what we view
as the smallest number of particles in a grain because the peak temperature
increases for smaller particle sizes and the probability of losing an atom
rises exponentially with the peak temperature.  If we consider a single
Mg$_2$SiO$_4$ molecule with $N=7$ atoms as the smallest grain, then we
find $E_1 \simeq 6$~eV ($9.0$~eV) if the grain starts from an equilibrium
temperature of $1500$~K ($1000$~K).  If, however, we view the smallest grain as
consisting of two such silicate units with $N=14$, then $E_1 \simeq 13.6$~eV
($21.0$~eV).  In either case, if the transient produces too many soft UV photons, small
grains will be destroyed by the radiation faster than they can grow.
If we define the absorption efficiency by $Q = Q'(a/\lambda)$, the rate at which such photons are
absorbed is
\begin{equation}
     t_\gamma^{-1} = {L_{phot} a^3 \over 4 r^2 h c } F(G=Q',E_0/kT_{phot},E_1/kT_{phot}) 
\end{equation}
where the function $F$ is the same as for the estimate of the number
of ionizing photons in Eqn.~\ref{eqn:specfunc} but with $G=Q'$ rather than $G=1/x$.
Dust can only grow once the evaporation rate is lower than the collisional
growth rate (Eqn.~\ref{eqn:dgrowth}), leading to a photoevaporation limit
on the mass loss rate for dust formation of
\begin{eqnarray}
    \dot{M} &> &{ v_w \over v_c } { L_{phot} a m_0 \over X_g h c } F \nonumber \\
            & = &3100 F  { v_w \over v_e } { \hbox{km/s} \over v_c } 
         { L_{phot} \over 10^6 L_\odot } \left( { 10^6 L_\odot\over L_* } \right)^{1/4} \left( { T_* \over 10^4~\hbox{K} } \right)
        \left( { M_* \over 20 M_\odot } \right)^{1/2}
    \left( { 0.005 \over X_g } \right) \left( { \hat{m}_0^4 N \over \hat{\rho} }\right)^{1/3} 
         { M_\odot \over \hbox{year}}.
\end{eqnarray}
The factor $(\hat{m}_0^4 N/\rho)^{1/3} \simeq 2$ for $N \simeq 7$.
The enormous difference between the photon and particle densities means 
that the possibility of dust formation is entirely controlled by the spectral
energy distribution of the transient and the value of $E_0$.  As with the
recombination limits, the differences between black bodies and actual
photospheres are crucial -- the limits on $\dot{M}$ for black bodies
are several orders of magnitude higher than those for the \cite{Castelli2003}
models.   As we see in Fig.~\ref{fig:dform1}, the photoevaporation
limit on $\dot{M}$ is a wall blocking dust formation in the quiescent
state independent of mass loss rate.  
Thus, dust can only form around hot stars if they do not appear to be hot when
observed from the dust formation radius.  These stars appear to have two means
of achieving this -- the S Doradus phase and (great) eruptions.

In the S Doradus phase, the luminosity of the star is little changed,
$L_{phot} \simeq T_*$, but the stars have cooler photospheric temperatures,
$T_{phot} \simeq 7000$~K.  For our model of the S Doradus phase, we adopt the more 
common view that the lower temperature is due to a true expansion of the
stellar photosphere rather than a ``pseudo-photosphere'' formed in a dense 
wind (see the discussion in \citealt{Vink2009}).  In the S Doradus state,
the stars have fairly high mass loss rates, $\dot{M} \sim 10^{-5}$--$10^{-4}M_\odot$, and fast
winds with $v_\infty \simeq v_e$, but they cannot be forming significant
amounts of dust even though they satisfy the condition on photospheric
temperature.  With a dust optical depth of
\begin{equation}   
  \tau_V = { \dot{M} \kappa_V \over 4 \pi v_w R_{form} }
       \simeq 3 \left( { \dot{M} \over 10^{-4} M_\odot/\hbox{year} } \right)
                \left( { \kappa_V \over 100~\hbox{cm}^2/\hbox{g} }\right)
                \left( { 10^{15}~\hbox{cm} \over R_{in} }\right) { v_e \over v_w }
\end{equation}
the stars would become bright, hot mid-IR sources and
some would be heavily enshrouded by their own dust, yet neither phenomenon
seems to be reported.  

As we show in the middle panels of Fig.~\ref{fig:dform1}, where we simply
set the apparent photospheric temperature to
$T_{phot} = \hbox{min}\left( T_*, 7000~\hbox{K} \right)$, the cooler
temperature is not sufficient to allow dust formation given the typical
mass loss rates.  First, the particle growth rates are too low.  Second,
the photosphere is still producing enough soft UV photons that the smallest
grains still tend to photo-evaporate faster than they can grow.  This
second limit is very sensitive to the minimum photon energy $E_0$ needed
to photo-evaporate a grain, but for $E_0 =10$~eV the limit is close to
the limit for any particle growth, at roughly $\dot{M} \gtorder 10^{-4}M_\odot$/year,
If we lower the minimum energy to $E_0=7.5$~eV, the required minimum mass loss rate
jumps enormously because we are counting photons on the rapidly falling
blue side of the spectrum.  Thus, while the precise limits are sensitive
to the exact choices of $T_{phot}$ and $E_0$, the combination of slow
growth and photoevaporation mean that dust cannot form in the S Doradus
phase.  

The final case we consider is a (giant) eruption where there is an
increase in the bolometric luminosity $L_{phot} > L_*$, the apparent
temperature is cooler, $T_{phot} < T_*$, as in the S Doradus phases,
and the mass loss rates are much higher, $\dot{M} > 10^{-2} M_\odot$/year.
For sufficiently dense winds, the dust formation region
sees a pseudo-photosphere created by the non-dust opacity of the
wind rather than the hot stellar photosphere.  \cite{Davidson1987} explains
this as a consequence of combining a dense wind with an opacity
law that is falling rapidly with temperature in this temperature
range.  Consider the Rosseland mean optical depth
\begin{equation}
       \tau_R(R) = \int_R^{R_3} \rho \kappa_R(\rho,T) dR
\end{equation}
looking inwards from the radius $R_3$ where the gas temperature 
is $1000$~K and dust formation may be possible at some interior
radius $R$.   If we combine the rapidly rising $\rho \propto r^{-2}$ density profile
of the wind with a temperature regime where the opacity rises
rapidly, then there will be a tendency to produce a 
pseudo-photosphere where $\tau_R(R)=1$ near that temperature.
We computed the temperature $T_w(R(\tau_R=1))$  at the radius
where $\tau_R=1$ using the 
solar composition opacity models of \cite{Helling2009}, 
our standard wind density profile and assuming a
temperature profile $T_w=T_*(R/R_*)^{-1/2}$.
This is not a self-consistent wind model, but the results
are insensitive to the assumptions because the opacity 
and optical depth increase so rapidly towards smaller
radii in the wind.  
Fig.~\ref{fig:dform1} shows the
consequences of using this ``pseudo-photospheric'' temperature in determining
the photo-evaporation limit rather than $T_*$, as well
as the contour where $T_w(R(\tau_R=1))=7000$~K.  

The limits now have two branches.  For small $\dot{M}$ or
low stellar temperatures, the wind is optically thin, the observed temperature
is simply the photospheric temperature and the photo-evaporation
limits are unchanged.  For high $\dot{M}$ and high temperatures, 
the wind becomes optically thick and the observed temperature
is of order $7000$~K with relatively weak dependencies on
$\dot{M}$ and $v_w$ because of the steep slope of the opacity,
as predicted by \cite{Davidson1987}. 
As expected from the arguments summarized by \cite{Vink2009}, 
the mass loss rates needed to form a pseudo-photosphere are 
higher than are typically found for the S Doradus phase. 
However, once $\dot{M} \gtorder 10^{-2.5}M_\odot$/year, the 
wind forms a pseudo-photosphere whose temperature slowly drops
with increasing mass loss rate, which makes the photevaporation 
limits less sensitive to $E_0$ than in our S Doradus model.
Note that in both the S Doradus and eruption models the photosphere 
must stay in its cool state long enough for the ejecta to reach the dust 
formation radius ($\sim 1$~year, Eqn.~\ref{eqn:rform}) if dust is to form.

Once dust forms, the size distribution then controls the opacity,
\begin{equation}
      \kappa_\lambda = { 3 X_d \over 4 \rho_{bulk} } { \qbar{a_{max}} \over a_{max} }, 
    \label{eqn:kappa}
\end{equation}
where the dimensionless function
\begin{equation}
       \qbar{a_{max}} = { a_{max} \int_0^{a_{max}} Q_\lambda(a) a^2 {dn \over da } da \over
                                 \int_0^{a_{max}} a^3 { dn \over da } da }
    \label{eqn:qbar}
\end{equation}
depends on the grain size distribution $dn/da$, the dimensionless (absorption or scattering)
cross section $Q_\lambda(a)$, and the fraction of the gas mass in condensed dust $X_d \leq X_g $.
The function $\qbar{a_{max}}$ is proportional to $a_{max}$ for very small grains, where $Q_\lambda \propto a$
and becomes constant for very large grains where $Q_\lambda$ becomes constant (e.g. \citealt{Draine1984}).
Thus, the ratio $\qbar{a_{max}}/a_{max}$
appearing in the opacity becomes constant for very small grains, decays as $a_{max}^{-1}$ for very
large grains and has a maximum at an intermediate size $a_{peak}$.  At V band
for a \cite{Mathis1977} size distribution $dn/da \propto a^{-3.5}$ with a range of $a_{max}/a_{min}=50$,
we find $a_{peak} \simeq 0.16\mu$m ($0.45\mu$m) with $\qbar{a_{peak}} \simeq 2.4$ ($\simeq 1.5$) and
$\qbar{a_{peak}}/a_{peak}=15.4\mu$m$^{-1}$ ($3.4\mu$m$^{-1}$) for graphitic (silicate) dust.
These estimates were made for the effective absorption optical depth $(\tau_{abs}(\tau_{abs}+\tau_{scat}))^{1/2}$
and lead to maximum visual opacities of
\begin{eqnarray}
         \kappa_{V,max}&\simeq &260 { X_d \over 0.005}~\hbox{cm}^2/\hbox{g} \qquad\hbox{graphitic} \nonumber \\
         \kappa_{V,max}&\simeq & 30 { X_d \over 0.005}~\hbox{cm}^2/\hbox{g} \qquad\hbox{silicate}.
         \label{eqn:maxkappa}
\end{eqnarray}
For $\tau_{abs}/\tau_{scat}$ the coefficients are $210$/$13$ and $130$/$75$ for graphitic/silicate dusts.
In general, however, the size dependence of the visual opacity is relatively weak.  If $a_{max}\gtorder 1\mu$m
the opacity begins to drop as $a_{max}^{-1}$ (Eqn.~\ref{eqn:amax}), but this requires 
$\dot{M} \gtorder M_\odot$/year, which no star seems to significantly exceed.
For very small grains, 
$\qbar{a_{max}}/a_{max}\rightarrow7.7\mu$m$^{-1}$ ($0.6\mu$m$^{-1}$) for graphitic (silicate) dust, 
so the opacity is only a factor of two (six) lower than the maximum opacity.  If the grains cannot
grow to moderate size, then the dust opacity will be significantly reduced.

\begin{figure}[p]
\centerline{\includegraphics[width=5.5in]{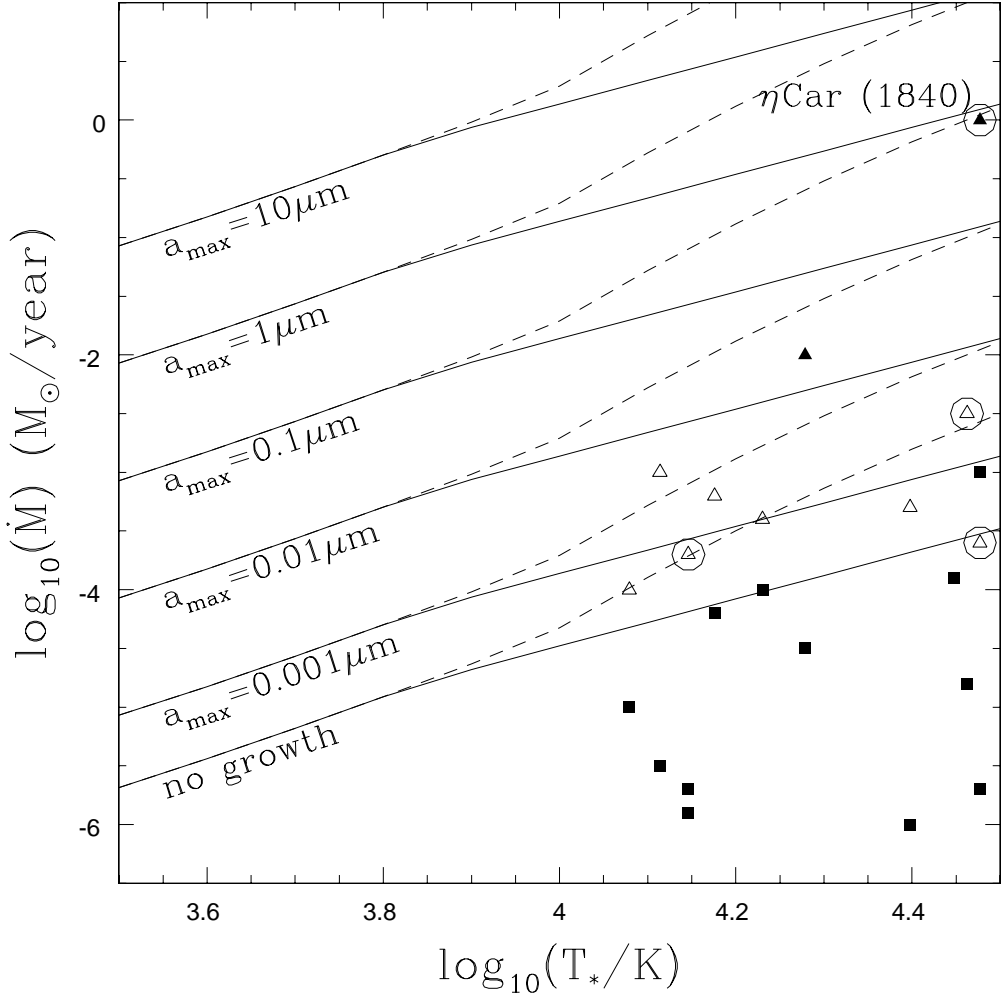}}
\caption{
  Mass loss rates needed to grow dust particles to radius $a_{max}$ (Eqn.~\ref{eqn:amax}).
  In computing the ratio of Planck factors $Q_{rat}$ we have either assumed the existence
  of a cooler photosphere $T_{phot}=\hbox{min}\left(T_*,7000~\hbox{K}\right)$ (solid) 
  or used $T_{phot}=T_*$ (dashed). 
  The triangles show the estimated mass loss rates during (great) eruptions based
  on either the observed duration (filled triangles, $\eta$~Car and P~Cyg) or durations
  estimated from the shell widths (open triangles).  We argue in the text that these latter
  estimates are gross underestimates of the mass loss rates in eruption.  
  The filled squares show the present day properties of the 
  systems in Table~1.  Objects in Table~1 noted as having exceptionally large grain sizes are 
  circled. {\it The temperatures are left fixed at the present day temperature estimates --
  in reality they were cooler during the eruption but we lack direct measurements.} 
  }
  \label{fig:size}
\end{figure}

\begin{figure}[p]
\centerline{\includegraphics[width=5.5in]{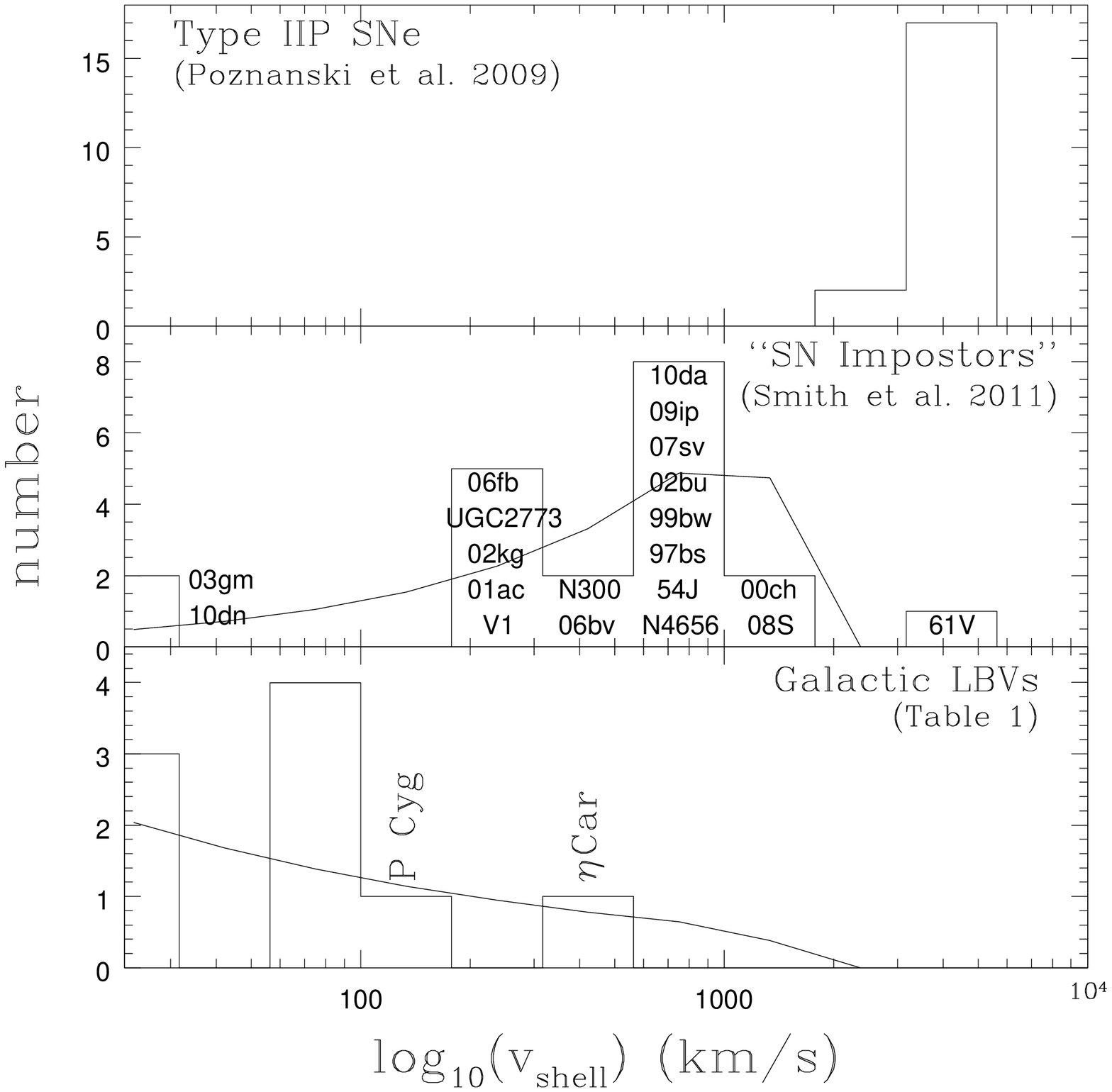}}
\caption{
  Asymptotic expansion velocities of Type~IIP~SNe (top), SN ``impostors'' (middle) and
  Galactic LBV shells (bottom).  The impostor velocities have been corrected
  for expansion out of the stellar potential well following Eqn.~\ref{eqn:vcor} in order
  to properly compare them to the LBV shells.  The general pattern of the 
  results is not sensitive to the details of this correction.  The SNe and
  the shells require no corrections because of their high velocities and
  ages, respectively.  The curves in the impostor and Galactic LBV panels
  show the expected distributions for the bins assuming a velocity-dependent rate $r(v_{shell}) \propto v_{shell}^{-1/3}$
  for $10~\hbox{km/s} < v_{shell} < 1500~\hbox{km/s}$ for the impostors and an
  observable lifetime $\propto 1/v_{shell}$ for the Galactic shells normalized
  to the numbers of objects in each panel and excluding SN~1961V.
  }
  \label{fig:vels}
\end{figure}

\section{Discussion}
\label{sec:discussion}

To summarize, the growth of dust particles in the ejecta of massive
stars is limited by particle growth rates and photo-evaporation by
soft, non-ionizing UV photons from the star.  The particle growth rate is the 
limiting factor for lower temperature stars ($T_* \ltorder 7000$~K), 
and photo-evaporation is the limiting factor for higher temperature 
stars ($T_* \gtorder 7000$~K).  While they are hot stars in their
quiescent state and cannot form dust in their winds, the LBVs have
cooler apparent temperatures, $T_{phot} \simeq 7000$~K, in their S 
Doradus phases and during (great) eruptions.  While they are cooler
in the S Doradus phase, their mass loss rates are also not high enough
for the grains to collisionally grow or to overcome the photoevaporation
of small grains by the remaining soft UV photons.  In this phase,
the mass loss rates are not high enough for the non-dust opacity 
of the wind to self-shield the dust formation region, so the cooler
temperature of the photosphere must be due to a true expansion of
the photosphere, as argued in the review by \cite{Vink2009}.  Only in (great)
eruptions with mass loss rates $\dot{M}\gtorder 10^{-2.5} M_\odot$/year
do these stars have the proper conditions for forming dust.  Moreover,
when the mass loss rates are this high,  the wind does form a  
``pseudo-photosphere'' with a temperature $T_{phot} \sim 7000$~K 
that shields the dust formation region from the soft UV emission 
of the true stellar photosphere.  This is the characteristic ``eruption'' temperature
of luminous blue variables (e.g. \citealt{Humphreys1994}) and it
is a consequence of the steep rise of the non-dust opacity with 
temperature in this regime (\citealt{Davidson1987}). 

This distinction between an expanded photosphere for the S Doradus state
and a ``pseudo-photosphere'' for the eruptions is also supported by the
momentum transfers needed for a radiatively accelerated wind.  Momentum
conservation requires $\dot{M} v_\infty \simeq \tau L/c$, where $\tau$
is the non-dust opacity source responsible for accelerating the wind
(see \citealt{Kudritzki2000}, \citealt{Puls2008}). 
For an asymptotic velocity of $v_\infty$,  the optical depth must be
\begin{equation}
       \tau \simeq 10^4 \left( { \dot{M} \over M_\odot/\hbox{year} }\right)
                   \left( { 10^6 L_\odot \over L_* } \right){ v_\infty \over v_{esc} }.
     \label{eqn:momentum}
\end{equation}
In the S Doradus phase, having $\dot{M} \ltorder 10^{-4}M_\odot$/year and
$v_\infty \simeq v_{esc}$ means that $\tau < 1$ and the wind cannot form
a pseudo-photosphere.  In the (great) eruptions with $\dot{M} \gtorder 10^{-2}M_\odot$/year,
the optical depth must be $\tau \gg 1$ and the wind must have a pseudo-photosphere
in order to be radiatively accelerated.  We should also note that once dust dust forms, 
it can be a significant source of acceleration because the Eddington factor for radiation
pressure on the dust,
\begin{equation}
        \Gamma_{dust} = { \kappa_{rp} L_{phot} \over 4 \pi c G M_* } 
                   \simeq 40 \left( { \kappa_{rp} \over 100~\hbox{cm}^2/\hbox{g} }\right)
                             \left( { L_{phot} \over 10^6 L_\odot } \right)
                             \left( { 20 M_\odot \over M_* } \right),
\end{equation}
is large.  Dust formation, as a large additional source of continuum opacity,
may help to address some of the problems in accelerating these heavy winds 
(e.g. \citealt{Owocki2004}).

The first important consequence of this close relationship between
dust formation and the need for very high mass loss rates is that 
the dust shells around luminous blue stars are formed exclusively
in great eruptions and so trace the history of these eruptions.  Given
a census of such dusty shells, their radii, expansion velocities, (dust) 
masses and optical depths in the Milky Way or other galaxies, 
it should be possible to reconstruct this dominant mass loss
mechanism for these massive stars.  A particularly interesting
diagnostic is the maximum grain size.  Where the total dust
mass or optical depth of a shell probes the total mass
lost in the eruption, the maximum grain size probes the mass
loss rate because, as shown in Fig.~\ref{fig:size}, the 
maximum grain size is proportional to the mass loss rate,
$a_{max} \propto \dot{M}$ (Eqn.~\ref{eqn:amax}).
Models of four of the Galactic shells appear to require 
$a_{max} \gtorder 1\mu$m (see Table~1) which strongly suggests 
$\dot{M} \gtorder 10^{-2}M_\odot$/year or possibly even higher.   

A second consequence is that the mass loss rates 
associated with most of the shells in Table~1 are grossly 
underestimated -- they are
too low to make any dust let alone super-sized grains.  These low estimates
of $\dot{M}$  come from the assumption that the duration of the
transient can be estimated from the radial thickness of the shell:
$\Delta t \simeq \Delta R/v_{shell} \simeq 10^4$years since
$\Delta R \simeq R_{shell} \simeq 1$~pc and $v_{shell} \simeq 70$~km/s.
This leads to an estimate of $\dot{M}=10^{-4} M_\odot$/year for
$M_{shell}=M_\odot$ that is not very different from many
of the present day winds which are not making dust, as 
illustrated in Fig.~\ref{fig:size}.  The flaw
here is that the observed spread in radius probably comes from
temporal and azimuthal variations in velocity rather than the
duration of the transient, just as we see in $\eta$ Carinae. 
This is proved by the simple geometric observation that all 
shells have comparable thickness ratios, which is the characteristic of a
spread in velocity: $\Delta R = \Delta v t$ and $R= v t$ so
$\Delta R/R = \Delta v/v$ is independent of time.  If it were 
due to the duration of the transient then $\Delta R = v \Delta t$
and $R= v t$ so $\Delta R/R = \Delta t/t$ and the shells
only appear geometrically thin as they become old. Roughly speaking,
for every shell with a 2:1 thickness ratio there should be one 
which is a filled sphere just finishing its eruption, and this is
not observed.  

It should be possible to determine the geometric structure of these shells
in some detail because many of the central stars are known to be significantly
variable (e.g. $\eta$ Carinae, see, e.g., \cite{Fernandez2009} for a full light curve,
or, e.g., \citealt{Martin2006} for spatially resolved data; 
AG~Car, \citealt{Groh2009}; IRAS~18576+3341, \citealt{Clark2009}).  You can
determine both the structure of the shell and obtain a geometric distance
to the source by mapping the time delay between the variability of the 
star and the echoes of the variability across the shell, by essentially
the same procedure as is used in reverberation mapping of quasars (see the review
by \citealt{Peterson1993})  or at a less involved level in 
studies of SN dust echoes (e.g. \citealt{Patat2005}).  This would
complement the proper motion measurements possible for some
systems (e.g. \cite{Smith2004} for $\eta$ Carinae).  The optimal
wavelength is probably on the blue side of the mid-IR peak,
at $10$-$20\mu$m to maximize the sensitivity to dust temperature
variations while minimizing the direct radiation from the star,
but scattered optical or near-IR emission is another possibility 
if the central star is faint enough to allow  imaging of the shell. 

With the exception of the Great Eruption of $\eta$ Carinae ($250$-$500$~km/s),
the typical expansion velocities of the Galactic shells are
only $50$-$100$~km/s (see Table~1, Fig.~\ref{fig:vels}).  As we argued 
earlier in \S2, these expansion velocities are unlikely to have been significantly
slowed by decelerations due to sweeping up the surrounding
interstellar medium and so must be associated with the ejection
mechanism.  The relatively low velocities of the Galactic shells mean that 
comparisons of the so-called ``SN impostors'' to LBV eruptions require
detailed examination.  Fig.~\ref{fig:vels} shows the expansion velocities
of a sample of normal Type~IIP SNe (\citealt{Poznanski2009}),
the Galactic eruptions from Table~1, and the SN ``impostors''
from \cite{Smith2011}.  The latter have been conservatively
corrected to an asymptotic expansion velocity at large radius by
\begin{equation}
     v_\infty^2 = v^2 - { 2 G M_* \over v t }
      \label{eqn:vcor}
\end{equation}
where we used $M_*=40M_\odot$ and $t=14$~days. 
While the corrections for some of the individual objects 
are sensitive to the choice of these parameters, the overall 
results are not.  In this recasting of the similar figure 
from \cite{Smith2011}, we see that almost 
none of the impostors have velocities similar to the Galactic 
shells.  We must, however, exercise care in comparing the velocity distributions
of impostors and Galactic shells in Fig.~\ref{fig:vels}
because slowly expanding shells are detectable for longer
periods of time, $t_{shell} \propto 1/v_{shell}$ 
(Eqn.~\ref{eqn:tshell}).  If the intrinsic rate
of eruptions with asymptotic velocities $v_{shell}$ is $r(v_{shell})$,
the number of observable Galactic shells is $\propto r(v_{shell})/v_{shell}$,
independent of any other consideration such as correlations
between $v_{shell}$ and $M_{shell}$ or completeness.  

Fig.~\ref{fig:vels} also shows a model for the velocity distributions
that is consistent with both samples.  We assumed the intrinsic rate
as a function of asymptotic velocity is a power law, 
$r(v_{shell}) \propto v_{shell}^\alpha$ with 
$v_{min} < v_{shell} < v_{max}$. We excluded SN~1961V since
it was probably an SN   (see \citealt{Kochanek2011}, \citealt{Smith2011}),
but otherwise ignored other ambiguities as to the nature of the impostor sample
(e.g. the very different physics of SN~2008S and the NGC~300-OT,
see \citealt{Kochanek2011b}).  The best fitting 
model has $\alpha \simeq -1/3$, $v_{min} \simeq 10$~km/s
and $v_{max} \simeq 1500$~km/s. Given the numbers of objects,
the uncertainties are large ($-0.75 < \alpha < 0.15$ for an
order of magnitude change in K-S test probabilities). 
If the true rate is $r(v_{shell}) \propto v_{shell}^\beta$,
then the difference can be interpreted as a velocity-dependent
completeness $c(v_{shell}) \propto v_{shell}^{\alpha-\beta}$. 
For example, if the true rate is independent of $v_{shell}$ ($\beta=0$),
then either the impostor sample is incomplete at low velocities or 
the Galactic sample is incomplete at high velocities.  
There can be additional biases created by asymmetries 
in the ejection velocities, since the early time velocities 
may represent the fastest expanding material while the late 
time shell emission may be dominated by the slowest moving 
material -- however, only some 50\% of the shells are strongly aspherical
(\citealt{Weis2011}) and the factor of $\sim 2$ asymmetry in
$\eta$~Carinae is not large enough to represent a significant bias.

We should note that velocity is not the only parameter in
which there is essentially no overlap between the Galactic
and impostor samples.  First, the eruption time scales of
the only two Galactic systems where they are known, $\eta$ Carinae 
and P~Cyg, are an order of magnitude (or more) longer than
those of almost all impostors (years to decades versus months,
see \citealt{Smith2011} for a summary), even though they are the
only Galactic systems with relatively high velocities.  Second, the 
ejected masses of the impostors almost certainly have to be
far smaller than the typical Galactic shell.  Assuming the
impostors are radiatively driven, energy conservation means
that the upper bounds on their ejected masses are 
$\ltorder 0.1 M_\odot$, while the typical Galactic shell
has a mass $\gtorder 1 M_\odot$.  Like the velocity distribution,
the mass differences can be driven by the lifetimes of the 
Galactic shells, $t_{shell} \propto M_{shell}^{-1/2}$
(Eqn.~\ref{eqn:tshell}).  Nonetheless, while the Galactic sources
and the impostors may be produced by a single process with
a broad parameter range (velocity, mass, time scale), the observed 
Galactic and extragalactic sources basically sample completely different regions of that parameter
space.   If these two populations are to be unified, then
the statistics and properties of the local systems need to
be systematically determined, and the completeness of the
extragalactic surveys need to be improved.

In addition to (probably) being the dominant mass loss mechanism for massive stars,
the eruptions may have a comparable importance to SNe as
a source of dust, particularly as a source of large grains due to their
favorable conditions for particle growth.
Suppose that the final pre-SN mass of
the stars undergoing eruptions is $M_0$ and that fraction $f$ of
the lost mass is in eruptions producing material with a dust-to-gas
ratio of $X_d$.  For SN to dominate the dust production by massive
stars, they must produce  
\begin{equation}
       M_{d,SN} \gtorder 0.09 f \left( {X_d\over 0.005} \right) 
             \left( { 40 M_\odot \over M_{erupt} } \right)^{0.35}
              \left[ 1 - 0.26 { M_0 \over M_{erupt}} + 
                         0.26 { M_0 \over M_{erupt}} \left( { M_e \over M_{up} } \right)^{0.35}
                       - \left( { M_e \over M_{up} } \right)^{0.35} \right]~M_\odot
\end{equation}  
of dust per SN.  Unlike \S1, we have allowed $M_{up}$ to be finite.
We know either from the arguments of \cite{Humphreys1984} and \cite{Smith2006a} or
our estimates of shell statistics in \S1 that $f \gtorder 0.5$, so eruptions
dominate the dust distribution unless $M_{d,SN} \gtorder 0.02 f M_\odot$
if we are conservative ($M_{erupt}=40M_\odot$, $M_0=10 M_\odot$, $M_{up}=100 M_\odot$)
or $M_{d,SN} \gtorder 0.06 f M_\odot$  if we are more liberal
($M_{erupt}=20M_\odot$, $M_0=5 M_\odot$, $M_{up}=300 M_\odot$).  Moreover,
the presence of a dense circumstellar medium may also enhance dust 
production by SNe (\citealt{Smith2008}).
While SN dust production rates are uncertain, few SNe show evidence for
producing this amount of dust (see the discussion in
\citealt{Matsuura2009}), although \cite{Matsuura2011} subsequently
reported the detection of $0.4$-$0.7M_\odot$ of dust associated with 
SN~1987A.  
 
There is also increasing evidence that such high mass loss phases are crucial to
understanding SNe on two levels.  The first problem is simply one of rates. For
the nominal parameters suggested by our discussion of the abundance of shells
in \S1, the rate of eruptions must be roughly the same as the rate of SNe.  
However, the rates of the faint Type~IIn SNe generally believed to correspond to 
such transients are significantly lower (e.g. \citealt{Li2011}).  This is 
consistent with the arguments in \cite{Thompson2009} and \cite{Horiuchi2011} that 
there is strong evidence for incompleteness in $M_V \ltorder -16$~mag transients.  
Many candidate eruptions are significantly fainter, with $-10.5 < M_V < -15$~mag (\citealt{Smith2011}), so 
the completeness of these surveys for eruptions is presumably still worse.
Local surveys need to find these fainter transients in order to make a
complete inventory.  The second issue is that there appears to be a
mismatch between massive star formation rates and SN rates of almost
a factor of two (\citealt{Horiuchi2011}).  If the underlying estimates of massive star
formation rates and SNe rates are correct, then either many of these fainter
transients need to be SNe or there must be a significant population of
failed SN (\citealt{Horiuchi2011}).  This is another facet of the need
to correctly classify the impostors in order to understand their statistics.   

The final issue we consider is that some supernova show evidence in their evolution
that they are interacting with the massive dense shells of material 
created by these eruptions.  The most dramatic examples are the hyperluminous
Type~IIn SNe (\citealt{Smith2007}), although these seem to be related
to low metallicity environments (\citealt{Kozlowski2010}, \citealt{Stoll2011}).  However,
\cite{Fox2011} argue that many Type~IIn SNe show evidence for CSM
interactions requiring the dense shells produced by eruptions.  
More generally, the existence of a dust echo from an SNe implies
an eruption within $10^3$-$10^4$~years whenever the progenitor is a hot star if
our theory that dust only forms in eruptions is correct.\footnote{This discussion
does not apply to dust echoes from the SNe of red supergiants where it is feasible to produce 
the dust in a slow, steady wind.}  In fact,
some hyperluminous SNe shows evidence for the presence of two shells --
an inner one to boost the total luminosity and an outer dusty shell (e.g.
\citealt{Smith2008b}, \citealt{Kozlowski2010}).  
The existence of any such correlation has dramatic implications for
the cause of stellar eruptions.  In order to produce any strong
SN interaction phenomena, the shell of material must have been produced within time 
$t_{int}=10^{1.5}$--$10^{2.5}$ years of the SN.  Suppose fraction
$f_{int} \simeq 10^{-2}$--$10^{-1}$ of SNe require the CSM densities of
eruptions, then the time period $\Delta t$ prior to the SNe over which
the ejections can be occurring is
\begin{equation}
    \Delta t = 0.1 N_{erupt} { t_{int} \over f_{int} } \left( { 40 M_\odot \over M_{erupt} }\right)^{1.35}  
           \simeq 10^{4} \left( { N_{erupt} \over 2 }\right) 
         \left( { t_{int} \over 300~\hbox{years} } \right) \left( { 0.01 \over f_{int} }\right)
            \left( { 40 M_\odot \over M_{erupt} }\right)^{1.35}~\hbox{years}.
         \label{eqn:snrelate}
\end{equation}
The existence of dust echoes leads to a similar conclusion -- while $t_{int}$ is larger,
their incidence in SNe $f_{int}$ is higher.
While this point has been made before in a qualitative sense (see, e.g. \citealt{Smith2010}, \citealt{Smith2011}), 
Eqn.~\ref{eqn:snrelate} makes it quantitatively clear how strong a constraint 
results from the existence of any such correlation.   Moreover, the parameters
chosen for the scaling in Eqn.~\ref{eqn:snrelate} may be significantly overestimating $\Delta t$. 
If any such correlation exists, massive shell ejections are forced to be associated with the very last phases of
massive star evolution, roughly to the onset of carbon burning, and this
suggests that the underlying driving mechanism is
post-carbon ignition nuclear burning instabilities (see the discussion in \citealt{Smith2007}).
Unfortunately, the only known 
mechanism of this kind, the pair instability SN (\citealt{Woosley2007}), requires very high
masses ($M_*\gtorder 100M_\odot$) and should not function at the
metallicities of any of these nearby examples.  There could still be a strong metallicity
effect because of the dependence of line-driven stellar winds on metallicity (see \citealt{Puls2008}).
As mass loss by normal winds becomes less efficient, stars may be more dependent on eruptions
for mass loss, although this begs the question of how eruptions might become more efficient
at lower metallicity.  This question could be addressed by the investigating the
statistics of LBV eruptions and shells as a function of environment.

\acknowledgements 

The author thanks K.~Davidson, R.~Humphreys,
M.~Pinsonneault, K.~Sellgren, N. Smith, K.Z.~Stanek, D.M.~Szczygiel, T.A. Thompson and B.E.~Wyslouzil
for comments and discussions.  C.S.K. is supported by National Science Foundation (NSF) grant AST-0908816

\begin{deluxetable}{lrrrrrrrrrl}
%\tabletypesize{\scriptsize}                        
\tablecaption{Summary of Galactic LBVs With Dusty Shells}
\tablewidth{0pt}
\tablehead{
\colhead{Object} &
  \colhead{$L_*$} &
  \colhead{$T_*$} &
  \colhead{$\dot{M}_{now}$} &
  \colhead{$v_{w,now}$} &
  \colhead{$R_{shell}$} &
  \colhead{$v_{shell}$ } &
  \colhead{$M_{shell}$}  &
  \colhead{$\dot{M}_{shell}$}  &
  \colhead{$a_{max}$} &
  \colhead{References} \\
 &
  \colhead{$L_\odot$} &
  \colhead{K} &
  \colhead{$M_\odot$/yr} &
  \colhead{km/s} &
  \colhead{pc} &
  \colhead{km/s} &
  \colhead{$M_\odot$}  &
  \colhead{$M_\odot$/yr}  &
  \colhead{$\mu$m} \\
}
\startdata
  $\eta$ Car (1840)    &$10^{6.7}$ &$30000$  &$10^{-3.0}$ &$500$        &$0.08$ &$250/500$ &$15$   &$10^{0.0}$   &$1$    &D97,S09 \\
  $\eta$ Car (1890)    &           &         &            &             &$0.03$ &$140/300$ &$0.1$  &$10^{-2.0}$  &       & \\
  Wray~17--96          &$10^{6.3}$ &$13000$  &$10^{-5.5}$ &$100$        &$1.0$  &          &$10$   &$10^{-3.0}$  &       &E02 \\
  AG Car               &$10^{6.2}$ &$29000$  &$10^{-4.8}$ &$110$        &$0.80$ &$70$      &$25$   &$10^{-2.5}$  &$10$   &S91,L94,V00 \\ 
  G79.29+0.46          &$10^{6.1}$ &$25000$  &$10^{-6.0}$ &$110$        &$1.8$  &$30$      &$14$   &$10^{-3.3}$  &       &W96,V00b,J10 \\ 
  G26.47+0.02          &$10^{6.0}$ &$17000$  &$10^{-4.0}$ &$\equiv 200$ &$2.3$  &          &$1.9$  &$10^{-3.4}$  &       &C03 \\
  P Cyg                &$10^{5.9}$ &$19000$  &$10^{-4.5}$ &$190$        &$0.07$ &$136$     &$0.1$  &$10^{-2.0}$  &       &N01, S06\\
  Wra~751              &$10^{5.8}$ &$30000$  &$10^{-5.7}$ &$\equiv 500$ &$0.34$ &$26$      &$1.7$  &$10^{-3.6}$  &$1$    &H91b,deW92,V00 \\
  %aka /Hen~3--591
  IRAS~18576           &$10^{5.8}$ &$15000$  &$10^{-4.2}$ &$160$        &$0.15$  &$70$     &$10$   &$10^{-3.2}$  &       &U01,U05,C09,B10 \\ 
  %aka /AFGL~2298
  % 70 km/s velocity is from Clark, Ueta models give outer radii, but they appear to be unconstrained and
  % simply set to be X times larger than the inner, using that outer radius then gives them very high masses
  % and old ages -- just peak emission from the later Buemi paper
  W~243                &$10^{5.8}$ &$18000$  &$10^{-5.4}$ &              &       &         &       &     &       &C04 \\ 
  Hen~3--519           &$10^{5.7}$ &$28000$  &$10^{-3.9}$ &$365$         &$1.1$  &$61$     &$0.66$ &     &       &S94 \\
  HR Car               &$10^{5.6}$ &$14000$  &$10^{-5.7}$ &$145$         &$0.3$  &$30/100$ &$3.0$  &     &       &H91a,L96,M02 \\
  G24.73+0.69          &$10^{5.6}$ &$12000$  &$10^{-5.0}$ &$\equiv 200$  &$1.6$  &         &$0.5$  &$10^{-4.0}$  &       &C03 \\
  HD~168625            &$10^{5.4}$ &$14000$  &$10^{-5.9}$ &$180$         &$0.48$ &$19$     &$0.25$ &$10^{-3.7}$  &$1$    &N96,P02,O03 \\ 
\enddata
\label{tab:summary}
\tablecomments{\hphantom{f}\hphantom{f} Luminosities and temperatures are from \cite{Smith2004b}.  
Other sources are 
B10 (\citealt{Buemi2010}),
C03 (\citealt{Clark2003}),
C04 (\citealt{Clark2004}),
C09 (\citealt{Clark2009}),
D97 (\citealt{Davidson1997}),
deW92 (\citealt{deWinter1992}),
E02 (\citealt{Egan2002}),
H91a (\citealt{Hutsemekers1991a}),
H91b (\citealt{Hutsemekers1991}),
J10 (\citealt{Jimenez2010}),
L96 (\citealt{Lamers1996}),
L94 (\citealt{Leitherer1994}),
M02 (\citealt{Machado2002}),
N01 (\citealt{Najarro2001}),
N96 (\citealt{Nota1996}),
O03 (\citealt{Ohara2003}),
P02 (\citealt{Pasquali2002}),
S91 (\citealt{Smith1991}),
S94 (\citealt{Smith1994}),
S06 (\citealt{Smith2006b}),
S09 (\citealt{Smith2009}), 
V00a (\citealt{Voors2000}),
V00b (\citealt{Voors2000b}),
U01 (\citealt{Ueta2001}),
U05 (\citealt{Umana2005}),
W96 (\citealt{Waters1996}).
The ejected mass estimates are generally derived from the mid-IR dust luminosities and assume a dust-to-gas ratio of $X_d=0.01$.
For $\eta$~Carinae the dust mass estimate agrees with estimates of the ejected gas mass (\citealt{Smith2007b}). The S06 estimate for P~Cyg
is a gas mass estimate.  The dust content of P~Cyg is uncertain, although there is a mid-IR excess (see S06).
  }
\end{deluxetable}

\end{document}